\documentstyle[prl,aps,epsfig]{revtex}

\tightenlines

\begin{document}

\draft

\title{S$_{17}$(0) Determined from the Coulomb Breakup of 83~MeV/nucleon
$^{8}$B}

\author {B.~Davids$^{1,2}$\cite{kvi}, D.W.~Anthony$^{1,3}$,
T.~Aumann$^{1}$\cite{gsi}, Sam~M.~Austin$^{1,2}$, T.~Baumann$^{1}$,
D.~Bazin$^{1}$, R.R.C.~Clement$^{1,2}$, C.N.~Davids$^{4}$, H.~Esbensen$^{4}$,
P.A.~Lofy$^{1,3}$, T.~Nakamura$^{1}$\cite{tit}, B.M.~Sherrill$^{1,2}$, and
J.~Yurkon$^{1}$}

\address {$^{1}$ National Superconducting Cyclotron Laboratory, Michigan State
University, East Lansing, Michigan 48824\\$^{2}$ Department of Physics and
Astronomy, Michigan State University, East Lansing, Michigan 48824\\$^{3}$
Department of Chemistry, Michigan State University, East Lansing, Michigan
48824\\$^{4}$ Physics Division, Argonne National Laboratory, Argonne, Illinois
60439}

\date{\today}

\maketitle

\begin{abstract} A kinematically complete measurement was made of the Coulomb
dissociation of $^{8}$B nuclei on a Pb target at 83 MeV/nucleon. The cross
section was measured at low relative energies in order to infer the
astrophysical S factor for the $^{7}$Be(\textit{p},$\gamma$)$^{8}$B reaction. A
first-order perturbation theory analysis of the reaction dynamics including
$E1$, $E2$, and $M1$ transitions was employed to extract the $E1$ strength
relevant to neutrino-producing reactions in the solar interior. By fitting the
measured cross section from $E_{rel}$~=~130~keV to 400~keV, we find
$S_{17}(0)$~=~17.8~$^{+1.4}_{-1.2}$~eV~b.\end{abstract}


The $\beta^{+}$ decay of $^{8}$B is the predominant source of high-energy solar
neutrinos. These neutrinos produce the most events in the chlorine radiochemical
detector at Homestake and the water and heavy water Cerenkov solar neutrino
detectors SuperKamiokande and SNO. In the Sun, $^{8}$B is produced via the
$^{7}$Be(\textit{p},$\gamma$)$^{8}$B reaction. Since 1964, the rate of this
reaction has been the most uncertain input to the calculated solar neutrino
fluxes, and the predicted event rates in solar neutrino detectors
\cite{bahcall}. Precise knowledge of this reaction rate is essential not only
for a detailed understanding of solar neutrino experiments, but also for
constraining fundamental properties of neutrinos themselves. Direct measurements
of the cross section are difficult because the target is radioactive, and the
cross section small.

Radiative capture cross sections are often characterized in terms of an
energy-dependent cross section factor, $S(E)~=~E~\sigma(E)~exp[2 \pi Z_{1} Z_{2}
e^{2} / (\hbar v)]$, where the $Z_{i}$ are the charges and $v$ the relative
velocity of the nuclei involved. Hammache {\em et al.} \cite{hammache} discuss
the discrepancies in the overall normalizations of the direct measurements of
the astrophysical S factor for the $^{7}$Be(\textit{p},$\gamma$)$^{8}$B
reaction, $S_{17}$. The disagreements among the direct measurements make an
independent approach desirable. Peripheral transfer reactions that yield
asymptotic normalization coefficients \cite{afshin} and Coulomb breakup
\cite{baur86,kiener,motobayashi,kikuchi97,kikuchi98,iwasa,baur96} permit the
extraction of S factors with different systematic uncertainties. In the Coulomb
breakup of $^{8}$B, a virtual photon emitted by a heavy target nucleus such as
Pb dissociates an incident $^{8}$B projectile into $^{7}$Be + \textit{p}. This
is the inverse of the radiative capture reaction. The two reaction rates are
related by the detailed balance theorem for photons of a given multipolarity.

As illustrated in \cite{iwasa}, there is a disagreement in the energy dependence
of the S factors extracted from the Coulomb dissociation experiments at RIKEN
and GSI. Furthermore, the radiative capture reaction proceeds almost exclusively
by $E1$ transitions at solar energies ($\approx$ 20 keV), but $E2$ and $M1$
transitions also play a role in Coulomb breakup for relative energies less than
1 MeV. $E2$ transitions are particularly important at low and intermediate beam
energies, while $M1$ transitions are most significant at high incident beam
energies. The contributions of these multipolarities to measured Coulomb
dissociation cross sections must be correctly accounted for in order to obtain
the $E1$ yield relevant to the production of $^{8}$B in the Sun. The size of the
$M1$ contribution at low relative energies can be gauged from the direct
measurement of the radiative capture cross section near the 0.64 MeV 1$^{+}$
resonance \cite{filippone}. The $E2$ contribution to Coulomb dissociation cross
sections was determined in an experiment by Davids {\em et al.} \cite{davids} in
which the longitudinal momentum distributions of $^{7}$Be fragments emitted in
the Coulomb dissociation of intermediate energy $^{8}$B projectiles on a Pb
target were measured. In that experiment, we observed an asymmetry in the
longitudinal momentum distribution of the emitted $^{7}$Be fragments
characteristic of interference between $E1$ and $E2$ transition amplitudes
\cite{esbensen}. In this Letter, we report an exclusive breakup measurement that
confirms the presence of $E2$ transitions in the Coulomb breakup, and
quantitatively account for the measured $E2$ contribution in inferring
$S_{17}$(0).

We made a kinematically complete measurement of the cross section for the
Coulomb dissociation of $^{8}$B on a Pb target at low relative energies. An
83~MeV/nucleon $^{8}$B beam delivered by the A1200 fragment separator
\cite{A1200} at the National Superconducting Cyclotron Lab impinged on a 47 mg
cm$^{-2}$ Pb target. The $^{8}$B beam intensity was approximately
10$^{4}$~s$^{-1}$; nearly 4 billion nuclei struck the target. A 1.5~T dipole
magnet separated the breakup fragments $^{7}$Be and \textit{p} from each other
and from the elastically scattered $^{8}$B nuclei, and dispersed the fragments
according to their momenta. Four multiwire drift chambers (MWDCs) were used to
measure the positions and angles of the breakup fragments after they passed
through the magnet. An array of 16 plastic scintillators was used for particle
identification. A thin scintillator at the exit of the A1200 provided continuous
measurements of the beam intensity. In conjunction with the plastic scintillator
array, it was also used to measure times-of-flight and to make intermittent beam
transmission and purity measurements. A stainless steel plate prevented most of
the direct $^{8}$B beam from reaching the detectors. Using the ion optics code
\textsc{cosy infinity} \cite{cosy}, we reconstructed the 4-momenta of the
breakup fragments from the measured positions in the four detectors and the
known magnetic field. The momentum calibration obtained from $^7$Be and proton
beams of known momenta was verified by checking that the fragment velocity
distributions were centered about the beam velocity.

The detection efficiency and experimental resolution were determined by means of
a Monte Carlo simulation, accounting for the beam emittance, energy loss and
multiple scattering in the target and detectors, and the detector position
resolution. The 1$\sigma$ relative energy resolution ranged e.g. from 100~keV at
$E_{rel}$~=~300 keV to 250~keV at $E_{rel}$~=~1.5 MeV. The 1$\sigma$ resolution
in the reconstructed angle of the dissociated $^{8}$B projectile was 4.5 mrad.
The dominant contribution to the experimental resolution was the position
resolution of the MWDCs. The simulation of the angular distribution of the
breakup fragments included both $E1$ and $E2$ transitions and anisotropic
breakup in the $^{8}$B center-of-mass system. Such an anisotropic angular
distribution was predicted by the model of Ref. \cite{esbensen}, and was
required to fit the longitudinal momentum distributions of protons measured in
the present experiment, and of $^{7}$Be fragments measured in \cite{davids}.
This point is addressed in detail below. The anisotropy is a consequence of
interference between $E1$ and $E2$ transition amplitudes.

The results of \cite{davids} imply that a proper theoretical description of a
$^{8}$B Coulomb breakup experiment must include $E2$ transitions. In Ref.
\cite{davids}, the analysis of the measured $^{7}$Be momentum distributions
assumed first-order perturbation theory using the point-like projectile
approximation for the Coulomb dissociation, and neglecting nuclear-induced
breakup. This was reasonable for the experimental conditions considered, namely,
for relatively small scattering angles of the $^{8}$B center-of-mass. The
analysis employed the $E1$ and $E2$ matrix elements predicted by the model of
Ref. \cite{esbensen}, scaled independently in order to reproduce the data. The
best fits for both incident beam energies, 44 and 81 MeV/nucleon, were obtained
when the ratio of the $E2$ and $E1$ matrix element scaling factors was 0.7. This
was incorrectly reported as the ratio of the scaling factors for the $E2$ and
$E1$ strength distributions; the correct value for this ratio is 0.7$^{2}$ =
0.49. As a consequence, the reported \cite{davids} ratio of $E2$ and $E1$ S
factors at $E_{rel}$ = 0.6 MeV should be replaced by 4.7$^{+2.0}_{-1.3} \times
10^{-4}$. The $E2$ strength extracted from the inclusive breakup measurement
\cite{davids} is a factor of 10 to 100 larger than the upper limits reported in
other experimental studies \cite{kikuchi97,iwasa}. However, it is only slightly
smaller than or in good agreement with recent theoretical calculations of $E2$
strength \cite{esbensen,typel,bennaceur,descouvement,barker}, and is consistent
with the measurement of \cite{guimaraes}. That the extracted experimental value
should be somewhat smaller than the theoretical values is consistent with the
idea that first-order perturbation theory overestimates the $E2$ contribution to
the cross section \cite{esbensen}.

In order to minimize the role of $E2$ transitions and possible nuclear
diffraction dissociation contributions to the breakup cross section measured in
this experiment, only events with $^{8}$B scattering angles of 1.8$^{\circ}$ or
less were analyzed, corresponding classically to an impact parameter of 30 fm.
Eikonal model \cite{bertulani98} and distorted-wave Born approximation
\cite{shyam} calculations find that nuclear-induced breakup is negligible up to
the grazing angle ($\approx$ 4$^{\circ}$), so the severe scattering angle cut
imposed here gives confidence that nuclear effects are small, and that the
point-like projectile approximation is valid. A first-order perturbation theory
analysis neglecting nuclear-induced breakup was employed to interpret the
results of this experiment. Such an approach is justified by the high beam
energy and the restricted angular range covered in the experiment. Higher-order
effects are most important at large scattering angles and low incident beam
energies \cite{baur96,esbensen}. Recent continuum-discretized coupled channels
calculations \cite{surrey} suggest that nuclear excitations account for less
than 4 \% of our measured breakup cross section below 500 keV, and that
higher-order electromagnetic processes have little effect on $d\sigma/dE_{rel}$
for the angles and energies covered in this experiment \cite{davids01}.

\begin{figure}\epsfig{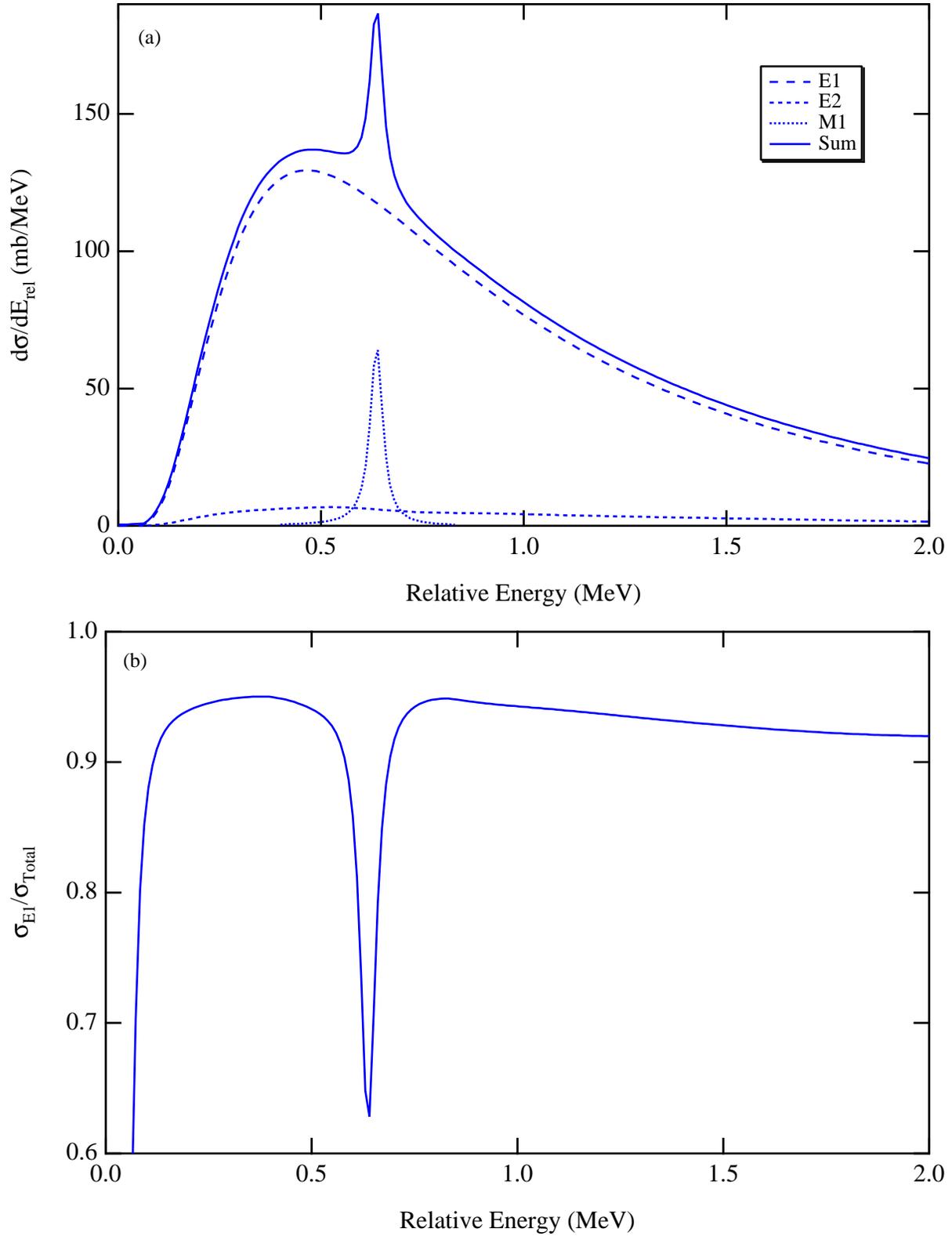} \caption{(a) Contributions of $E1$, $E2$,
and $M1$ transitions to the cross section for the Coulomb dissociation of 83
MeV/nucleon $^{8}$B on Pb with $^{8}$B scattering angles of 1.8$^{\circ}$ or
less in 1st-order perturbation theory. $M1$, $E1$, and $E2$ cross sections are
calculated as described in the text. (b) Fraction of the calculated cross
section for the Coulomb dissociation of 83 MeV/nucleon $^{8}$B on Pb with
$^{8}$B scattering angles $\leq 1.8^{\circ}$ (b~$\geq$~30~fm) accounted for by
$E1$ transitions in 1st-order perturbation theory.} \label{fig1} \end{figure}

A particular strength of our analysis is that it includes all of the relevant
electromagnetic multipole contributions, $E1$, $E2$, and $M1$. The procedure was
the following. The $E1$ and $E2$ contributions were calculated using the
structure model of Ref. \cite{esbensen}, quenching the $E2$ matrix elements as
discussed earlier. The $M1$ contribution at the 0.64 MeV 1$^{+}$ resonance was
calculated by folding the measured $M1$ S factor \cite{filippone} with the $M1$
photon spectrum calculated in 1st-order perturbation theory \cite{bertulani88}.
The contributions of the different multipolarities and their sum is shown in
Fig.\ \ref{fig1} (a). By requiring $\Theta_{^{8}B}$~$\leq~1.8^{\circ}$ and
E$_{rel}~\geq$~130~keV, we have ensured the dominance of $E1$ transitions.
Except for a narrow range surrounding the $M1$ resonance, $E1$ transitions
represent over 90\% of the cross section in first-order perturbation theory.
Fig.\ \ref{fig1}(b) shows the fraction of the measured cross section accounted
for by $E1$ transitions in the present experiment.

\begin{figure} \epsfig{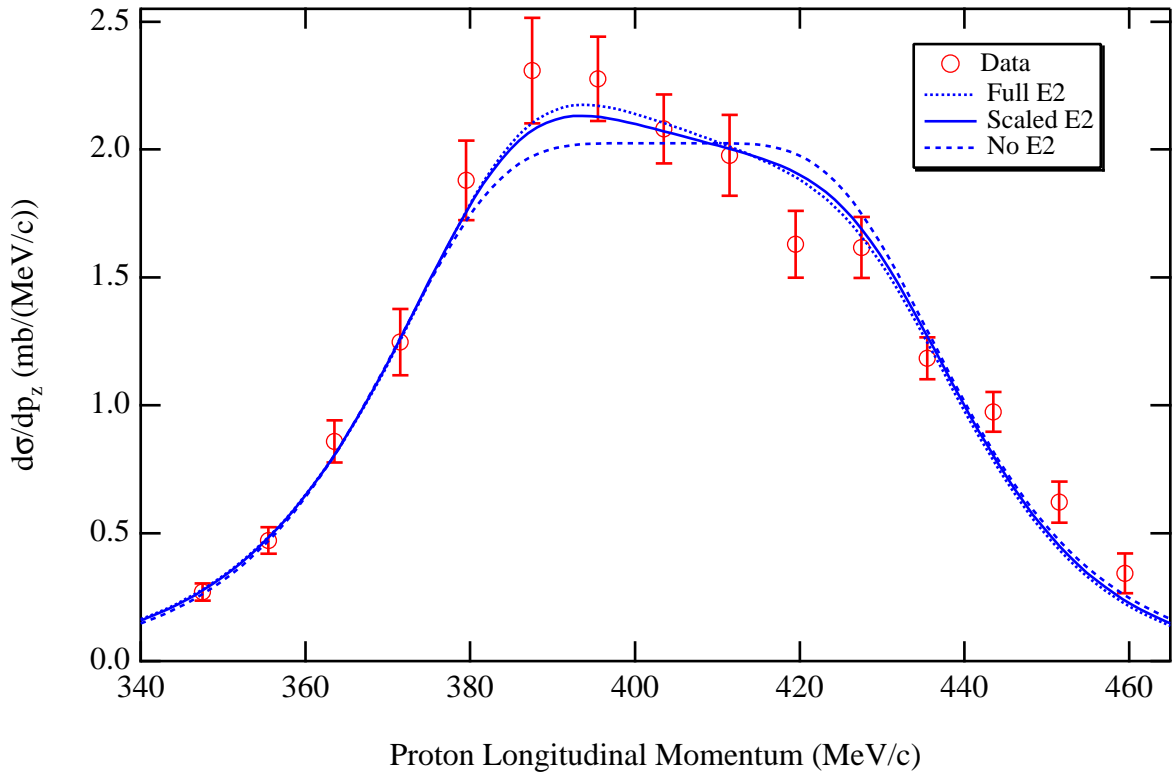} \vspace{5mm} \caption{Longitudinal
momentum distribution of protons emitted in the Coulomb dissociation of 83
MeV/nucleon $^{8}$B on Pb with $^{8}$B scattering angles of 1.8$^{\circ}$ or
less. The curves are 1st-order perturbation theory calculations using the model
of Ref. \protect\cite{esbensen} modified as described in the text, convoluted
with the experimental resolution. The error bars indicate the size of the
relative uncertainties.} \label{fig2} \end{figure}

The measured longitudinal momentum distribution of protons emitted in the
Coulomb breakup of 83~MeV/nucleon $^{8}$B on Pb with $^{8}$B scattering angles
of 1.8$^{\circ}$ or less is shown in Fig.\ \ref{fig2}. The 1$\sigma$ proton
momentum resolution was estimated from the simulation to be 4~MeV/c. Since the
statistical significance of these data is less than that of the inclusive
measurement reported in Ref. \cite{davids}, we shall not use them to extract the
$E2$ strength. Nevertheless, the asymmetry of this distribution is manifest.
Also shown in the figure are calculations done with the model of Ref.
\cite{esbensen}, one with the full $E2$ strength, one with the $E2$ matrix
elements scaled as described above, and another with no $E2$ matrix elements.
The asymmetry observed in \cite{davids}, taken together with momentum
conservation, implies that the proton longitudinal momentum distribution must
have a complementary asymmetry. We observed such an asymmetry for the first time
in this measurement, confirming the presence of $E2$ transitions in the Coulomb
breakup of $^{8}$B.

In analyzing the measured decay energy spectrum, we convoluted the sum of the
calculated $E1$, $E2$, and $M1$ contributions with the experimental resolution,
and scaled the magnitude of the $E1$+$E2$ contribution in order to minimize
$\chi^{2}$ for the data points within two energy intervals, 130~keV to 2~MeV,
and 130 to 400~keV. The factor by which the $E1$+$E2$ contribution was
multiplied will be referred to as the normalization factor. The data above 2 MeV
were excluded from the fit due to the presence of a 3$^{+}$ resonance at 2.2 MeV
that was not included in the theoretical calculation, and because the statistics
there are poor. At energies below 100 keV, our calculations show that the $E2$
component dominates, so these data were also excluded from the fit. A correction
to the data for the feeding of the 429 keV excited state of $^{7}$Be was made
using the results of \cite{kikuchi97}. This correction is small, ranging from
less than a percent at the lowest relative energies to about 10\% around 2 MeV.

The best-fit normalization factor obtained for the data between 130~keV and
2~MeV with this procedure was 1.00~$^{+0.02}_{-0.06}$. The 1$\sigma$ error
includes energy-dependent contributions from statistics, momentum and angular
acceptance, detector efficiency, and the $^{7}$Be excited state feeding
correction. The various sources of systematic uncertainties include beam
intensity (1\%), target thickness (2.6\%), momentum calibration (4.2\%), and the
theoretical uncertainty (5.6\%), resulting in a total systematic uncertainty of
7.5\%. The theoretical uncertainty includes contributions from the size of the
$E2$ component (2.5\%) and from the extrapolation to zero energy (5\%). Hence
the result of the perturbation theory analysis of data from 130~keV to 2~MeV is
$S_{17}$(0)~=~19.1~$^{+1.5}_{-1.8}$~eV~b.

A more reliable result can be obtained by analyzing a smaller relative energy
range. Jennings {\em et al.} \cite{jennings} point out that nuclear structure
uncertainties increase significantly above $E_{rel}$~=~400~keV. In order to
minimize this model dependence, we also fit only the data between 130~keV and
400~keV. The theoretical extrapolation uncertainty is only 1\% for this energy
range \cite{jennings}. The best-fit normalization factor for these data was
0.93~$^{+0.05}_{-0.04}$, resulting in $S_{17}(0)$~=~17.8~$^{+1.4}_{-1.2}$~eV~b,
with all sources of uncertainty added in quadrature. This result is consistent
with the value extracted from all the data up to 2~MeV, implying that the simple
potential model of Ref. \cite{esbensen} describes the physics well even at large
relative energies, within the uncertainties. The data and the best-fit 1st-order
perturbation theory calculations for all the data between 130~keV and 2~MeV, and
for the data from 130~to~400~keV, convoluted with the experimental resolution,
are shown in Fig.\ \ref{fig3}.

The present result is in good agreement with three of the capture measurements
\cite{hammache,filippone,vaughn}, and with the RIKEN (18.9 $\pm$ 1.8 eV b) and
GSI (20.6 $\pm$ 1.2 $\pm$ 1.0 eV b) Coulomb breakup measurements
\cite{kikuchi98,iwasa}. It is also in excellent agreement with the results of
asymptotic normalization coefficient determinations (17.3 $\pm$ 1.8 eV b)
\cite{afshin,anc}. Although the results agree within the errors, the $E1$
strength found here is about 15\% smaller than reported in the GSI Coulomb
breakup measurement \cite{iwasa}. This might be ascribed to the neglect of $E2$
transitions in the analysis of the GSI measurement. The fraction of the breakup
cross section attributable to $E2$ transitions depends on the energies and
angles covered. In the present measurement, the experimental conditions were
tailored to minimize the role of $E2$ transitions. The GSI measurement probed
smaller impact parameters, implying in first-order perturbation theory a
$\sigma_{E2}/\sigma_{E1}$ ratio about 4 times larger than in this measurement
throughout the relative energy range used to extract the S factor. Since the
$E2$ contribution to the present measurement is about 5\%, this could account
for the difference between the extracted $E1$ components. Similarly, the S
factor inferred from the RIKEN Coulomb breakup measurement \cite{kikuchi98} must
be reduced by 4-15\% \cite{motobayashi00} in order to account for the $E2$
contribution in 1st-order perturbation theory.

In summary, a kinematically complete measurement of the Coulomb dissociation of
83 MeV/nucleon $^{8}$B on a Pb target was carried out using a dipole magnet to
separate the breakup fragments from the beam. The Coulomb breakup cross section
was measured at low relative energies and small $^{8}$B scattering angles in
order to infer the astrophysical S factor for the
$^{7}$Be(\textit{p},$\gamma$)$^{8}$B reaction with minimal complications from
nuclear-induced breakup, $E2$ transitions, and higher-order electromagnetic
effects. A first-order perturbation theory description of the reaction that
included $E1$, $E2$, and $M1$ transitions, and a simple, single-particle $^{8}$B
structure model were used to interpret the measurement. The longitudinal
momentum distribution of the emitted protons was measured and found to be
asymmetric, consistent with our prior inclusive measurement of the $^{7}$Be
fragments, confirming the role of $E2$ transitions in the Coulomb breakup.
Although we obtained data below 100~keV, they were excluded from the analysis
because $E2$ transitions dominate at these energies. In order to minimize the
theoretical uncertainties, the $E1$ strength in the Coulomb breakup was
extracted from 130~to~400~keV, yielding
$S_{17}(0)$~=~17.8~$^{+1.4}_{-1.2}$~eV~b. Having for the first time properly
accounted for the $E2$ component, the dominant theoretical uncertainty in
$^{8}$B Coulomb breakup measurements, we have shown that direct radiative
capture, Coulomb breakup and asymptotic normalization coefficient determinations
give consistent values of $S_{17}(0)$.

This work was supported by the U.S. National Science Foundation; two of us
(C.N.D. and H.E.) were supported by the U.S. Department of Energy, Nuclear
Physics Division, under Contract No. W-31-109-ENG-38.

\begin{figure}\epsfig{file=fig3.epsf} \vspace{5mm} \caption{Measured cross
section for the Coulomb dissociation of 83 MeV/nucleon $^{8}$B on Pb with
$^{8}$B scattering angles $\leq 1.8^{\circ}$. Only relative errors are shown.
Also depicted are the best-fit 1st-order perturbation theory calculations for
the data between 130~keV and 2~MeV, and for the data between 130~keV and
400~keV, convoluted with the experimental resolution. The data point at 64~keV
was excluded from the fit because of a large $E2$ contribution.} \label{fig3}
\end{figure}

\end{document}